\documentstyle[12pt]{article}

\begin{document}

\title{{\Large {\bf Ricci and matter collineations of locally rotationally
symmetric space-times}}}
\author{Michael Tsamparlis and Pantelis S. Apostolopoulos \and {\it \ {\small %
Department of Physics, Section of Astrophysics-Astronomy-Mechanics, }} \and 
{\it {\small University of Athens, Zografos 15783, Athens, Greece} }}
\maketitle

\begin{abstract}
A new method is presented for the determination of Ricci Collineations (RC)
and Matter Collineations (MC) of a given spacetime, in the cases where the
Ricci tensor and the energy momentum tensor are non-degenerate and have a
similar form with the metric. This method reduces the problem of finding the
RCs and the MCs to that of determining the KVs whereas at the same time uses
already known results on the motions of the metric. We employ this method to
determine all hypersurface homogeneous locally rotationally symmetric
spacetimes, which admit proper RCs and MCs. We also give the corresponding
collineation vectors. These results conclude a long due open problem, which
has been considered many times partially in the literature.
\end{abstract}


\section{Introduction}

The field equations of General Relativity are highly non-linear pdfs and
their solution requires simplifying assumptions in the form of additional
conditions/constraints. There are many ways to impose simplifying
assumptions on the metric. These assumptions must satisfy various general
rules one of them being the requirement that they must be consistent with
the symmetry group of the metric and the geometric structures on the
spacetime manifold. One class of assumptions, which satisfy the above
demand, are the {\em collineations} or {\em geometric symmetries}.

A general collineation is defined ${\cal L}_{{\bf \xi }}A=\Phi $ where $A $
is any of the quantities $g_{ab}, \Gamma _{bc}^a, R_{ab}, R_{bcd}^a$ and
geometric objects constructed by them and $\Phi $ is a tensor with the same
index symmetries as $A$. There are many types of collineations defined by
the various forms of the tensors $A,\Phi .$ For example $A _{ab}=g_{ab}$
and $\Phi _{ab}=2\psi g_{ab}$ define a Conformal Killing vector (CKV), which
specializes to a Special Conformal Killing vector (SCKV) when $\psi _{;ab}=0$%
, to a Homothetic vector field (HVF) when $\psi =$constant and to a Killing
vector (KV) when $\psi =0.$ When $A _{ab}=R_{ab}$ and $\Phi _{ab}=2\psi
R_{ab} $ the symmetry vector $\xi ^a$ is called a Ricci Conformal
Collineation (RCC) and specializes to a Ricci Collineation (RC) when $%
\Phi _{ab}=0$. When $A _{ab}=T_{ab}$ and $\Phi _{ab}=2\psi T_{ab},$ where $T_{ab}$
is the energy momentum tensor, the vector $\xi ^a$ is called a Matter
Conformal Collineation (MCC) and specializes to a Matter collineation (MC)
when $\Phi _{ab}=0$. The function $\psi $ in the case of CKVs is called the
conformal factor and in the case of conformal collineations the {\em %
conformal function}.

It is well known that two different collineations are not in general
equivalent. For example a KV is a RC or a MC but the opposite does not hold.
Collineations have been classified by means of their relative properness in 
\cite{Katzin-Levine-Davies,Katzin-Levine}. From this classification it is
seen that the basic collineation is the KVs.

The role of the KVs (or symmetries) is to restrict (possibly with additional
assumptions) the general form of the metric. This results in a reduction of
the number of the independent field equations and (as a rule) in the
simplification of their study. We should note that there are well known
spacetime metrics which do not have KVs \cite{Szekeres}. This does not
exclude the possibility that they can admit higher collineations.

The role of a higher collineation is to supply the field equations with
additional equations, which are the equations defining the collineation.
These later equations involve the metric functions and the components of the
vector field defining the collineation.

Obviously the constraints imposed by additional collineations do not
guarantee that they will lead to a solution of the field equations. However
if they do then these solutions are compatible with the general structure of
the metric and the geometry resulting from it.

The standard method to deal with the augmented system of field equations is
the direct solution of the system of partial differential equations. As
expected, this procedure is in general difficult and, in many cases, it has
the defect that one can loose solutions, especially those occurring as
particular cases. As a result people have tried to find indirect methods of
solution, which relay more on differential geometry and less on the solution
of partial differential equations.

The purpose of this paper is twofold:\newline
(a) to present a ``practical'' method, which reduces (whenever this is
possible) the computation of the RCs and the MCs of a given metric to the
computation of KVs and\newline
(b) To apply this method and determine all hypersurface orthogonal locally
rotationally symmetric (LRS) spacetime metrics, which admit proper MCs and
proper RCs. We recall that a RC/MC is proper if it is not a KV or a HKV or a
SCKV.

A first partial exposition of the method has been given previously in \cite
{Apostol-Tsamp6} and independently in \cite{Camci and Barnes} and has been
applied in the determination of all Robertson-Walker metrics, which admit
proper RCs and MCs.

The proposed method applies only when the Ricci tensor $R_{ab}$ and the
energy momentum tensor $T_{ab}$ are non-degenerate and when the form of $%
R_{ab}$, $T_{ab}$ (equivalently $G_{ab}$) is similar to the form of the
metric. Let us assume that this is the case. Then one can consider on the
spacetime manifold the three metric elements\footnote{%
It is possible the Ricci tensor of a spacetime metric to be degenerate but
the Einstein tensor to be non-degenerate. See for example the stiff perfect
fluid ($\gamma=2$) LRS spacetime discussed in example 2 of section \ref
{SubSecExamples2}.} $ds_R^2=R_{ab}dx^adx^b,ds_T^2=T_{ab}dx^adx^b$ and $%
ds_g^2=g_{ab}dx^adx^b$, which have in general different signature, all
signatures being possible for the first two. Each of these `metrics` has a
symmetry group and because the ($C^\infty $) KVs of the metric are ($%
C^\infty $) RCs and ($C^\infty $) MCs these groups have a common subgroup.
But this subgroup is (as a rule) the main factor, which defines the general
form of the metric. Therefore it is logical to expect that \ $g_{ab,}R_{ab}$
and $T_{ab}$ will have a `common form', provided the definition of the
metric does not involve other assumptions besides KVs, for example discrete
symmetries. However this is not in general known or easy to prove and the
best way to make sure that this is the case it is to compute directly the
tensors $R_{ab},T_{ab}$ and see if their form is or is not similar to the
form of the metric.

If the answer is positive then it is possible to consider the `generic' line
element $ds^2=K_{ab}dx^adx^b$, which reduces to the three line elements $%
ds_g^2,ds_R^2,ds_T^2$ for appropriate choices of the coefficients $K_{ab}$.
The gains from this consideration are twofold.

a. If one solves Killing's equations for the generic metric then one has
found simultaneously the KVs of the metric, the RCs and the MCs. That is,
the problem of finding the RCs and MCs is reduced to that of calculating the
KVs.

b. The generic metric is possible to have all signatures. Therefore
Killing's equations will have to be solved for all possible signatures. In
case the KVs of the metric are known then only the signatures $(+,+,+,+)$
and $(+,+,-,-)$ need to be considered.

In case the form of the tensors $R_{ab},T_{ab}$ is different from the form
of the metric $g_{ab}$ the introduction of $K_{ab}$ makes no sense and one
has to follow the standard way, i.e. solve the pdfs resulting from the
constraint. However we note, that in many cases one can still introduce $%
K_{ab}$ for the tensors $R_{ab},T_{ab}$ only, and apply the same method.

It is useful to comment briefly when $R_{ab},T_{ab}$ are degenerate. In this
case it has been shown \cite{Hall-Roy-Vaz} that, in general, there are
infinitely many RCs and MCs, which must be found by the solution of the
relevant pdfs. However, the RCs in the degenerate case are not as useful as
the ones of the non-degenerate case. Indeed the assumption of degeneracy of $%
R_{ab},T_{ab}$ leads to differential equation(s), which fix the metric
functions up to arbitrary constants of integration. Hence the form of the
Ricci or the Matter tensor can be determined making the constraint imposed
by the RC/MC redundant. For example if $T_{ab}$ is degenerate it has been
shown \cite{Carot-daCosta-Vaz} that the only interesting case is when rank $%
T_{ab}=1$ that is, a null Einstein-Maxwell field or a dust fluid. In most
applications (including the LRS case as will be seen in the next section)
the corresponding metrics are known and there is no need for an extra
assumption to obtain the solution of the field equations.

The structure of the paper is as follows. In section 2 we consider the three
possible classes of LRS metrics. In sections 3,4 and 5 we determine the KVs
of the generic metric (whenever it can be defined) for all possible
signatures and give explicitly the proper RCs and the proper MCs. We also
discuss examples which show how our general results can be applied in
practice. Finally in section 6 we conclude the paper.

\section{Hypersurface homogeneous LRS spacetimes}

Hypersurface homogeneous spacetimes which are locally rotationally symmetric
(LRS spacetimes) contain many well known and important solutions of Einstein
field equations and have been studied extensively in the literature \cite
{Ellis1,Stewart-Ellis,Ellis-MacCallum}. They admit a group of motions $G_4$
acting multiply transitively on three dimensional orbits spacelike ($S_3)$
or timelike ($T_3)$ the isotropy group being a spatial rotation. It is well
known that there are three families of metrics describing these spacetimes 
\cite{Ellis1,Stewart-Ellis}: 
\begin{eqnarray}
ds^2 &=&\varepsilon [dt^2-A^2(t)dx^2]+B^2(t)\left[ dy^2+\Sigma
^2(y,k)dz^2\right]  \label{sx1.1} \\
ds^2 &=&\varepsilon \left\{ dt^2-A^2(t)\left[ dx+\Lambda (y,k)dz\right]
^2\right\} +B^2(t)\left[ dy^2+\Sigma ^2(y,k)dz^2\right]  \label{sx1.2} \\
ds^2 &=&\varepsilon [dt^2-A^2(t)dx^2]+B^2(t)e^{2x}(dy^2+dz^2)  \label{sx1.3}
\end{eqnarray}
where $\varepsilon =\pm 1,\Sigma (y,k)=\sin y,\sinh y,y$ and $\Lambda
(y,k)=\cos y,\cosh y,y^2$ for $k=1,-1,0$ respectively. (The factor $%
\varepsilon =\pm 1$ essentially distinguishes between the ''static'' and the
''nonstatic'' cases as it can be seen by interchanging the co-ordinates $t,x$%
). According to the classification made by Ellis \cite{Ellis1} the metrics (%
\ref{sx1.2}) with $\varepsilon =1$ are class $I$ LRS metrics, the metrics (%
\ref{sx1.1}) and (\ref{sx1.3}) are class $II$ and finally class $III$ are
the metrics (\ref{sx1.2}) with $\varepsilon =-1$.

As we have already remarked, the solution of the field equations for these
metrics is possible only in special cases, that is, when the metric is
required to satisfy additional constraints. In this paper we consider the
extra constraint to be the requirement that the LRS metric admits a proper
RC or a proper MC. To compute the subset of the LRS metrics selected by this
constraint we apply - when it is possible - the method of generic metric
described above. As it will be shown the method applies to the LRS metrics (%
\ref{sx1.1}) and (\ref{sx1.2}) and does not always applies to the metric (%
\ref{sx1.3}), in which case we have to work differently.

\section{Ricci and Matter Collineations of the LRS Ellis class II metrics
(1.1)}

The LRS metrics (\ref{sx1.1}) admit the isometry group $G_{4}$ consisting of
the four KVs $\partial _{x},{\bf X}_{\mu }$ $(\mu =1,2,3)$: 
\begin{equation}
{\bf X}_{\mu }=(\delta _{\mu }^{1}\cos z+\delta _{\mu }^{2}\sin z)\partial
_{y}-\left[ (\ln \Sigma )_{,y}(\delta _{\mu }^{1}\sin z-\delta _{\mu
}^{2}\cos z)-\delta _{\mu }^{3}\right] \partial _{z}  \label{sx2.1}
\end{equation}
acting on 3D spacelike or timelike orbits ($\epsilon =-1,1$ respectively).

We only consider the case $\epsilon =-1$ (3D spacelike orbits) because the
results for $\epsilon =1$ follow from the interchange of the coordinates $%
t,x $ (the Ricci and Matter tensor are identical up to a minus sign). Indeed
the Ricci tensor $R_{ab}$ and the Einstein tensor $G_{ab}$ (i.e. the energy
momentum tensor) are computed to be:

\begin{equation}
R_{ab}=\epsilon \cdot diag\left\{ \frac{2\ddot{B}A+\ddot{A}B}{AB},-\frac{%
A\left( \ddot{A}B+2\dot{B}\dot{A}\right) }B,-\frac{A\left( \ddot{B}%
B+B^2+k\right) +\dot{B}\dot{A}B}A\left[ 1,\Sigma ^2(y,k)\right] \right\}
\label{sx2.1a}
\end{equation}
\begin{equation}
G_{ab}=\epsilon \cdot diag\left\{ -\frac{A\left( \dot{B}^2+k\right) +2\dot{B}%
\dot{A}B}{AB^2},\frac{A^2\left( 2\ddot{B}B+\dot{B}^2+k\right) }{B^2},\frac{%
B\left( \ddot{A}B+\ddot{B}A+\dot{B}\dot{A}\right) }A\left[ 1,\Sigma
^2(y,k)\right] \right\} .  \label{sx2.1b}
\end{equation}
We note that the form of $R_{ab}$ and $G_{ab}$ is similar to that of the
metric, therefore for the metrics (\ref{sx1.1}) we can consider the
``generic'' metric: 
\begin{equation}
ds^2=K_0dt^2+K_1dx^2+K_2\left[ dy^2+\Sigma ^2(y,k)dz^2\right]  \label{sx2.2}
\end{equation}
which reduces to the metrics $ds_g^2,ds_R^2,ds_T^2$ when $%
K_a=\{g_a,R_a,G_a\} $ where:

\begin{equation}
g_a=\{-1,A^2,B^2,B^2\}  \label{sx2.2a}
\end{equation}
\begin{equation}
R_a=\left\{ -\frac{2\ddot{B}A+\ddot{A}B}{AB},\frac{A\left( \ddot{A}B+2\dot{B}%
\dot{A}\right) }B,\frac{A\left( \ddot{B}B+B^2+k\right) +\dot{B}\dot{A}B}%
A\left[ 1,1\right] \right\}  \label{sx2.2b}
\end{equation}
\begin{equation}
G_a=\left\{ \frac{A\left( \dot{B}^2+k\right) +2\dot{B}\dot{A}B}{AB^2},-\frac{%
A^2\left( 2\ddot{B}B+\dot{B}^2+k\right) }{B^2},-\frac{B\left( \ddot{A}B+%
\ddot{B}A+\dot{B}\dot{A}\right) }A\left[ 1,1\right] \right\} .
\label{sx2.2c}
\end{equation}
In order to compute the KVs of the generic metric we apply the
transformation: 
\begin{equation}
d\tilde{\tau}=\left| K_0\right| ^{1/2}dt  \label{sx2.3}
\end{equation}
so that: 
\begin{equation}
ds^2=\varepsilon _1(K_0)d\tilde{\tau}^2+K_1dx^2+K_2\left[ dy^2+\Sigma
^2(y,k)dz^2\right]  \label{sx2.4}
\end{equation}
where $\varepsilon _1(K_0)$ is the sign of the component $K_0$.

The KVs ${\bf X}$ of $ds^2$ are computed from the solution of Killing's
equations ${\cal L}_{{\bf X}}K_{ab}=0.$ Because the generic metric can have
all possible signatures we have to consider three cases:

\begin{itemize}
\item  Lorentzian case $(-1,1,1,1)$.

\item  Euclidean case $(1,1,1,1)$.

\item  The case $(-1,-1,1,1)$.
\end{itemize}

Killing's equations for the case of the Lorentzian signature have been
solved in \cite{Apostol-Tsamp5} where it has been shown that there are only
two possible cases to consider, i.e. non conformally flat metrics and the
conformally flat metrics. This conclusion holds for the other two cases
because it is independent from the signature of the generic metric.

The results of the (typical) calculations for the Lorentzian cases are
collected in Table 1 and for the non-Lorentzian cases in Table 2. Concerning
the explanation of Tables 1,2 we have the following. Classes $A_1-A_7$ refer
to the non-conformally flat cases and classes $B_1$ to $B_8$ to the
conformally flat cases. The columns $K_1,K_2$ give the functional forms of
the generic metric components in order the collineation(s) {${\bf X}$} to be
admitted. $\dim {\cal C}$ is the dimension of the full isometry group of the
``generic'' metric element $ds^2$ including the four vectors (\ref{sx2.1}).
Last column gives the expression of the KV(s) in terms of the coordinates
and the parameters entering the metric functions. It is worth noting that if 
{\em we interchange }$t,x$ in the expressions for the vector fields we
obtain the KVs for the {\em static case}.

Some of the collineations in Tables 1 and 2 have been found previously by
various authors (see for example \cite{Camci-Yavuz,Camci-Sharif} and
references cited therein).

The RCs and the MCs we give in Tables 1,2 are proper because they cannot be
reduced to the extra KV of the homogeneous or to the HVF of the self similar
corresponding spacetime. By demanding this reduction we have found for each
vector the values of the parameters, which should be excluded. \pagebreak 

\begin{center}
{\bf {\small Table 1}.}{\small \ KVs of the metrics (\ref{sx1.1}) for the
case }$(K_0K_1K_2)<0${\small , that is Lorentzian signature. }$k${\small \
is the curvature of the 2-space }$y,z${\small . }$\dim {\cal C}${\small \ is
the dimension of the full symmetry algebra of the generic metric }$ds^2$%
{\small .}

\begin{tabular}{|l|l|l|l|l|l|}
\hline
{\bf Class} & ${\bf k}$ & $K_1$ & $K_2$ & $\dim {\cal C}$ & ${\bf X}$ \\ 
\hline
$A_1$ & $0$ & $\pm c^2e^{-2\tilde{\tau}/\alpha _1c}$ & $\pm c^2e^{-2\tilde{%
\tau}/c}$ & {\bf 5} & $
\begin{tabular}{l}
$\alpha _1c\partial _{\tilde{\tau}}+x\partial _x+\alpha _1y\partial _y$ \\ 
$c\neq -\sqrt{2+\frac 1{a_1^2}},\;c\neq \frac{sign(b)}{2a_1\sqrt{1+2a_1^2-3b}%
}$%
\end{tabular}
$ \\ \hline
$A_2$ & $\pm 1$ & $\pm c_1^2c_2^2$ & $\pm c_2^2$ & {\bf 6} & $
\begin{array}{c}
\partial _{\tilde{\tau}} \\ 
c_1c_2x\partial _{\tilde{\tau}}+\frac{\tilde{\tau}}{c_1c_2}\partial _x
\end{array}
$ \\ \hline
$A_3$ & $0,\pm 1$ & $\pm c_1^2e^{\frac{2\tilde{\tau}}{ac_2}}$ & $\pm c_2^2$
& {\bf 6} & $
\begin{array}{c}
-ac_2\partial _{\tilde{\tau}}+x\partial _x \\ 
2ac_2x\partial _{\tilde{\tau}}-\left( x^2+\frac{a^2c_2^2}{c_1^2}e^{-\frac{2%
\tilde{\tau}}{ac_2}}\right) \partial _x
\end{array}
$ \\ \hline
$A_4$ & $0,\pm 1$ & $\pm c^2\cosh ^2\frac{\tilde{\tau}}{ac}$ & $\pm c^2$ & 
{\bf 6} & $
\begin{array}{c}
c\sin \frac xa\partial _{\tilde{\tau}}+\tanh \frac{\tilde{\tau}}{ac}\cos
\frac xa\partial _x \\ 
c\cos \frac xa\partial _{\tilde{\tau}}-\tanh \frac{\tilde{\tau}}{ac}\sin
\frac xa\partial _x
\end{array}
\;\mid ac\mid \neq 1$ \\ \hline
$A_5$ & $0,\pm 1$ & $\pm c^2\sinh ^2\frac{\tilde{\tau}}{ac}$ & $\pm c^2$ & 
{\bf 6} & $
\begin{array}{c}
c\sinh \frac xa\partial _{\tilde{\tau}}-\coth \frac{\tilde{\tau}}{ac}\cosh
\frac xa\partial _x \\ 
c\cosh \frac xa\partial _{\tilde{\tau}}-\coth \frac{\tilde{\tau}}{ac}\sinh
\frac xa\partial _x
\end{array}
\;\mid ac\mid \neq 1$ \\ \hline
$A_6$ & $0,\pm 1$ & $\pm c^2\cos ^2\frac{\tilde{\tau}}{ac}$ & $\pm c^2$ & 
{\bf 6} & $
\begin{array}{c}
c\sinh \frac xa\partial _{\tilde{\tau}}+\tan \frac{\tilde{\tau}}{ac}\cosh
\frac xa\partial _x \\ 
c\cosh \frac xa\partial _{\tilde{\tau}}+\tan \frac{\tilde{\tau}}{ac}\sinh
\frac xa\partial _x
\end{array}
\;\mid ac\mid \neq 1$ \\ \hline
$A_7$ & $\pm 1$ & $\pm \tilde{\tau}^2$ & $\pm c^2$ & {\bf 6} & $
\begin{array}{c}
\cosh x\partial _{\tilde{\tau}}-\tilde{\tau}^{-1}\sinh x\partial _x \\ 
\sinh x\partial _{\tilde{\tau}}-\tilde{\tau}^{-1}\cosh x\partial _x
\end{array}
$ \\ \hline
\end{tabular}
\newpage

{\bf {\small Table 1 (continued)}.}{\small \ KVs of the metrics (\ref{sx1.1}%
) for the case }$(K_0K_1K_2)<0${\small , that is Lorentzian signature. }$k$%
{\small \ is the curvature of the 2-space }$y,z${\small . }$\dim {\cal C}$%
{\small \ is the dimension of the full symmetry algebra of the generic
metric }$ds^2${\small .}

\begin{tabular}{|l|l|l|l|l|l|}
\hline
{\bf Class} & ${\bf k}$ & $K_1$ & $K_2$ & $\dim {\cal C}$ & ${\bf X}$ \\ 
\hline
$B_1$ & $1$ & $\pm c_1^2c^2$ & $\pm c^2\cosh ^2\frac{\tilde{\tau}}c$ & {\bf 7%
} & $
\begin{array}{c}
{\bf X}_{\mu +\nu +3}=-f_{(\mu )}\left[ f_{(\nu )}^{\prime }\right] _{,%
\tilde{\tau}}\left( c\cosh \frac{\tilde{\tau}}c\right) ^2\partial _{\tilde{%
\tau}}+ \\ 
+\frac{f_{(\mu )}\left[ f_{(\nu )}^{\prime }\right] _{,x}}{c_1^2\tanh ^2%
\frac{\tilde{\tau}}c}\partial _x-f_{(\nu )}^{\prime }\left[ f_{(\mu
)}\right] _{,y}\partial _y-\frac{f_{(\nu )}^{\prime }\left[ f_{(\mu
)}\right] _{,z}}{\sin ^2y}\partial _z
\end{array}
$ \\ \hline
$B_2$ & $-1$ & $\pm c_1^2c^2$ & $\pm c^2\sinh ^2\frac{\tilde{\tau}}c$ & {\bf %
7} & $
\begin{array}{c}
{\bf X}_{\mu +\nu +3}=-f_{(\mu )}\left[ f_{(\nu )}^{\prime }\right] _{,%
\tilde{\tau}}\left( c\sinh \frac{\tilde{\tau}}c\right) ^2\partial _{\tilde{%
\tau}}+ \\ 
+\frac{f_{(\mu )}\left[ f_{(\nu )}^{\prime }\right] _{,x}}{c_1^2\coth ^2%
\frac{\tilde{\tau}}c}\partial _x-f_{(\nu )}^{\prime }\left[ f_{(\mu
)}\right] _{,y}\partial _y-\frac{f_{(\nu )}^{\prime }\left[ f_{(\mu
)}\right] _{,z}}{\sinh ^2y}\partial _z
\end{array}
$ \\ \hline
$B_3$ & $-1$ & $\pm c_1^2c^2$ & $\pm c^2\sin ^2\frac{\tilde{\tau}}c$ & {\bf 7%
} & $
\begin{array}{c}
{\bf X}_{\mu +\nu +3}=-f_{(\mu )}\cdot \left[ f_{(\nu )}^{\prime }\right] _{,%
\tilde{\tau}}\left( c\sin \frac{\tilde{\tau}}c\right) ^2\partial _{\tilde{%
\tau}}+ \\ 
+\frac{f_{(\mu )}\left[ f_{(\nu )}^{\prime }\right] _{,x}}{c_1^2\cot ^2\frac{%
\tilde{\tau}}c}\partial _x-f_{(\nu )}^{\prime }\left[ f_{(\mu )}\right]
_{,y}\partial _y-\frac{f_{(\nu )}^{\prime }\left[ f_{(\mu )}\right] _{,z}}{%
\sinh ^2y}\partial _z
\end{array}
$ \\ \hline
$B_4$ & $1$ & $\pm c_1^2c^2\sinh ^2\frac{\tilde{\tau}}c$ & $\pm c^2\cosh ^2%
\frac{\tilde{\tau}}c$ & {\bf 10} & $
\begin{array}{c}
{\bf X}_{2(\mu +1)+\nu }=-f_{(\mu )}\left[ f_{(\nu )}^{\prime }\right] _{,%
\tilde{\tau}}\left( c\cosh \frac{\tilde{\tau}}c\right) ^2\partial _{\tilde{%
\tau}}+ \\ 
+\frac{f_{(\mu )}\left[ f_{(\nu )}^{\prime }\right] _{,x}}{c_1^2\tanh ^2%
\frac{\tilde{\tau}}c}\partial _x-f_{(\nu )}^{\prime }\left[ f_{(\mu
)}\right] _{,y}\partial _y-\frac{f_{(\nu )}^{\prime }\left[ f_{(\mu
)}\right] _{,z}}{\sin ^2y}\partial _z
\end{array}
$ \\ \hline
$B_5$ & $-1$ & $\pm c_1^2c^2\cosh ^2\frac{\tilde{\tau}}c$ & $\pm c^2\sinh ^2%
\frac{\tilde{\tau}}c$ & {\bf 10} & $
\begin{array}{c}
{\bf X}_{2(\mu +1)+\nu }=-f_{(\mu )}\left[ f_{(\nu )}^{\prime }\right] _{,%
\tilde{\tau}}\left( c\sinh \frac{\tilde{\tau}}c\right) ^2\partial _{\tilde{%
\tau}}+ \\ 
+\frac{f_{(\mu )}\left[ f_{(\nu )}^{\prime }\right] _{,x}}{c_1^2\coth ^2%
\frac{\tilde{\tau}}c}\partial _x-f_{(\nu )}^{\prime }\left[ f_{(\mu
)}\right] _{,y}\partial _y-\frac{f_{(\nu )}^{\prime }\left[ f_{(\mu
)}\right] _{,z}}{\sinh ^2y}\partial _z
\end{array}
$ \\ \hline
$B_6$ & $-1$ & $\pm c_1^2c^2\cos ^2\frac{\tilde{\tau}}c$ & $\pm c^2\sin ^2%
\frac{\tilde{\tau}}c$ & {\bf 10} & $
\begin{array}{c}
{\bf X}_{2(\mu +1)+\nu }=-f_{(\mu )}\left[ f_{(\nu )}^{\prime }\right] _{,%
\tilde{\tau}}\left( c\sin \frac{\tilde{\tau}}c\right) ^2\partial _{\tilde{%
\tau}}+ \\ 
+\frac{f_{(\mu )}\left[ f_{(\nu )}^{\prime }\right] _{,x}}{c_1^2\cot ^2\frac{%
\tilde{\tau}}c}\partial _x-f_{(\nu )}^{\prime }\left[ f_{(\mu )}\right]
_{,y}\partial _y-\frac{f_{(\nu )}^{\prime }\left[ f_{(\mu )}\right] _{,z}}{%
\sinh ^2y}\partial _z
\end{array}
$ \\ \hline
$B_7$ & $0$ & $\pm c_1^2$ & $\pm c_2^2$ & {\bf 10} & $
\begin{array}{c}
{\bf X}_{2(\mu +1)+\nu }=-c_2f_{(\mu )}\left[ f_{(\nu )}^{\prime }\right] _{,%
\tilde{\tau}}\partial _{\tilde{\tau}}+\frac{c_2f_{(\mu )}\left[ f_{(\nu
)}^{\prime }\right] _{,x}}{c_1^2}\partial _x- \\ 
-\frac{f_{(\nu )}^{\prime }\left[ f_{(\mu )}\right] _{,y}}{c_2}\partial _y-%
\frac{f_{(\nu )}^{\prime }\left[ f_{(\mu )}\right] _{,z}}{y^2c_2}\partial _z
\\ 
{\bf X}_9=\partial _{\tilde{\tau}} \\ 
{\bf X}_{10}=c_1x\partial _{\tilde{\tau}}+\frac{\tilde{\tau}}{c_1}\partial _x
\end{array}
$ \\ \hline
$B_8$ & $0$ & $\pm c_1^2\tilde{\tau}^2$ & $\pm c_2^2$ & {\bf 10} & $
\begin{array}{c}
{\bf X}_{2(\mu +1)+\nu }=-c_2f_{(\mu )}f_{(\nu )}^{\prime }\partial _{\tilde{%
\tau}}+\frac{c_2f_{(\mu )}\left[ f_{(\nu )}^{\prime }\right] _{,x}}{c_1^2%
\tilde{\tau}}\partial _x- \\ 
-\frac{\tilde{\tau}f_{(\nu )}^{\prime }\left[ f_{(\mu )}\right] _{,y}}{c_2}%
\partial _y-\frac{\tilde{\tau}f_{(\nu )}^{\prime }\left[ f_{(\mu )}\right]
_{,z}}{y^2c_2}\partial _z \\ 
{\bf X}_9=\cosh c_1x\partial _{\tilde{\tau}}-\frac 1{c_1\tilde{\tau}}\sinh
c_1x\partial _x \\ 
{\bf X}_{10}=\sinh c_1x\partial _{\tilde{\tau}}-\frac 1{c_1\tilde{\tau}%
}\cosh c_1x\partial _x
\end{array}
$ \\ \hline
\end{tabular}
\newpage

{\bf {\small Table 2}. }{\small KVs of the metrics (\ref{sx1.1}) for the
case }$(K_0K_1K_2)>0${\small , that is the signatures }$(+,+,+,+)${\small \
and }$(-,-,+,+)${\small . }$k${\small \ is the curvature of the 2-space }$%
y,z ${\small . }$\dim {\cal C}${\small \ is the dimension of the full
symmetry algebra of the generic metric }$ds^2${\small . }

\begin{tabular}{|l|l|l|l|l|l|}
\hline
{\bf Class} & ${\bf k}$ & $K_1$ & $K_2$ & $\dim {\cal C}$ & ${\bf X}$ \\ 
\hline
$A_1$ & $0$ & $\pm c^2e^{-2\tilde{\tau}/\alpha _1c}$ & $\pm c^2e^{-2\tilde{%
\tau}/c}$ & {\bf 5} & $\alpha _1c\partial _{\tilde{\tau}}+x\partial
_x+\alpha _1y\partial _y$ \\ \hline
$A_2$ & $\pm 1$ & $\pm c_1^2c_2^2$ & $\pm c_2^2$ & {\bf 6} & $
\begin{array}{c}
\partial _{\tilde{\tau}} \\ 
c_1c_2x\partial _{\tilde{\tau}}-\frac{\tilde{\tau}}{c_1c_2}\partial _x
\end{array}
$ \\ \hline
$A_3$ & $0,\pm 1$ & $\pm c_1^2e^{\frac{2\tilde{\tau}}{ac_2}}$ & $\pm c_2^2$
& {\bf 6} & $
\begin{array}{c}
-ac_2\partial _{\tilde{\tau}}+x\partial _x \\ 
2ac_2x\partial _{\tilde{\tau}}-\left( x^2-\frac{a^2c_2^2}{c_1^2}e^{-\frac{2%
\tilde{\tau}}{ac_2}}\right) \partial _x
\end{array}
$ \\ \hline
$A_4$ & $0,\pm 1$ & $\pm c^2\cos ^2\frac{\tilde{\tau}}{ac}$ & $\pm c^2$ & 
{\bf 6} & $
\begin{array}{c}
c\sin \frac xa\partial _{\tilde{\tau}}-\tan \frac{\tilde{\tau}}{ac}\cos
\frac xa\partial _x \\ 
c\cos \frac xa\partial _{\tilde{\tau}}+\tan \frac{\tilde{\tau}}{ac}\sin
\frac xa\partial _x
\end{array}
$ \\ \hline
$A_5$ & $0,\pm 1$ & $\pm c^2\sinh ^2\frac{\tilde{\tau}}{ac}$ & $\pm c^2$ & 
{\bf 6} & $
\begin{array}{c}
c\sin \frac xa\partial _{\tilde{\tau}}+\coth \frac{\tilde{\tau}}{ac}\cos
\frac xa\partial _x \\ 
c\cos \frac xa\partial _{\tilde{\tau}}-\coth \frac{\tilde{\tau}}{ac}\sin
\frac xa\partial _x
\end{array}
$ \\ \hline
$A_6$ & $0,\pm 1$ & $\pm c^2\cosh ^2\frac{\tilde{\tau}}{ac}$ & $\pm c^2$ & 
{\bf 6} & $
\begin{array}{c}
c\sinh \frac xa\partial _{\tilde{\tau}}-\tanh \frac{\tilde{\tau}}{ac}\cosh
\frac xa\partial _x \\ 
c\cosh \frac xa\partial _{\tilde{\tau}}-\tanh \frac{\tilde{\tau}}{ac}\sinh
\frac xa\partial _x
\end{array}
$ \\ \hline
$A_7$ & $\pm 1$ & $\pm \tilde{\tau}^2$ & $\pm c^2$ & {\bf 6} & $
\begin{array}{c}
\cos x\partial _{\tilde{\tau}}-\tilde{\tau}^{-1}\sin x\partial _x \\ 
\sin x\partial _{\tilde{\tau}}+\tilde{\tau}^{-1}\cos x\partial _x
\end{array}
$ \\ \hline
\end{tabular}
\newpage

{\bf {\small Table 2 (continued)} }{\small KVs of the metrics (\ref{sx1.1})
for the case }$(K_0K_1K_2)>0${\small , that is the signatures }$(+,+,+,+)$%
{\small \ and }$(-,-,+,+)${\small . }$k${\small \ is the curvature of the
2-space }$y,z${\small . }$\dim {\cal C}${\small \ is the dimension of the
full symmetry algebra of the generic metric }$ds^2${\small .}

\begin{tabular}{|l|l|l|l|l|l|}
\hline
{\bf Class} & ${\bf k}$ & $K_1$ & $K_2$ & $\dim {\cal C}$ & ${\bf X}$ \\ 
\hline
$B_1$ & $1$ & $\pm c_1^2c^2$ & $\pm c^2\cos ^2\frac{\tilde{\tau}}c$ & {\bf 7}
& $
\begin{array}{c}
{\bf X}_{\mu +\nu +3}=f_{(\mu )}\left[ f_{(\nu )}^{\prime }\right] _{,\tilde{%
\tau}}\left( c\cos \frac{\tilde{\tau}}c\right) ^2\partial _{\tilde{\tau}}+
\\ 
+\frac{f_{(\mu )}\left[ f_{(\nu )}^{\prime }\right] _{,x}}{c_1^2\tan ^2\frac{%
\tilde{\tau}}c}\partial _x-f_{(\nu )}^{\prime }\left[ f_{(\mu )}\right]
_{,y}\partial _y-\frac{f_{(\nu )}^{\prime }\left[ f_{(\mu )}\right] _{,z}}{%
\sin ^2y}\partial _z
\end{array}
$ \\ \hline
$B_2$ & $-1$ & $\pm c_1^2c^2$ & $\pm c^2\cosh ^2\frac{\tilde{\tau}}c$ & {\bf %
7} & $
\begin{array}{c}
{\bf X}_{\mu +\nu +3}=f_{(\mu )}\left[ f_{(\nu )}^{\prime }\right] _{,\tilde{%
\tau}}\left( c\cosh \frac{\tilde{\tau}}c\right) ^2\partial _{\tilde{\tau}}+
\\ 
+\frac{f_{(\mu )}\left[ f_{(\nu )}^{\prime }\right] _{,x}}{c_1^2\tanh ^2%
\frac{\tilde{\tau}}c}\partial _x-f_{(\nu )}^{\prime }\left[ f_{(\mu
)}\right] _{,y}\partial _y-\frac{f_{(\nu )}^{\prime }\left[ f_{(\mu
)}\right] _{,z}}{\sinh ^2y}\partial _z
\end{array}
$ \\ \hline
$B_3$ & $1$ & $\pm c_1^2c^2$ & $\pm c^2\sinh ^2\frac{\tilde{\tau}}c$ & {\bf 7%
} & $
\begin{array}{c}
{\bf X}_{\mu +\nu +3}=f_{(\mu )}\left[ f_{(\nu )}^{\prime }\right] _{,\tilde{%
\tau}}\left( c\sinh \frac{\tilde{\tau}}c\right) ^2\partial _{\tilde{\tau}}+
\\ 
+\frac{f_{(\mu )}\left[ f_{(\nu )}^{\prime }\right] _{,x}}{c_1^2\coth ^2%
\frac{\tilde{\tau}}c}\partial _x-f_{(\nu )}^{\prime }\left[ f_{(\mu
)}\right] _{,y}\partial _y-\frac{f_{(\nu )}^{\prime }\left[ f_{(\mu
)}\right] _{,z}}{\sin ^2y}\partial _z
\end{array}
$ \\ \hline
$B_4$ & $1$ & $\pm c_1^2c^2\sin ^2\frac{\tilde{\tau}}c$ & $\pm c^2\cos ^2%
\frac{\tilde{\tau}}c$ & {\bf 10} & $
\begin{array}{c}
{\bf X}_{2(\mu +1)+\nu }=f_{(\mu )}\left[ f_{(\nu )}^{\prime }\right] _{,%
\tilde{\tau}}\left( c\cos \frac{\tilde{\tau}}c\right) ^2\partial _{\tilde{%
\tau}}+ \\ 
+\frac{f_{(\mu )}\left[ f_{(\nu )}^{\prime }\right] _{,x}}{c_1^2\tan ^2\frac{%
\tilde{\tau}}c}\partial _x-f_{(\nu )}^{\prime }\left[ f_{(\mu )}\right]
_{,y}\partial _y-\frac{f_{(\nu )}^{\prime }\left[ f_{(\mu )}\right] _{,z}}{%
\sin ^2y}\partial _z
\end{array}
$ \\ \hline
$B_5$ & $-1$ & $\pm c_1^2c^2\sinh ^2\frac{\tilde{\tau}}c$ & $\pm c^2\cosh ^2%
\frac{\tilde{\tau}}c$ & {\bf 10} & $
\begin{array}{c}
{\bf X}_{2(\mu +1)+\nu }=f_{(\mu )}\left[ f_{(\nu )}^{\prime }\right] _{,%
\tilde{\tau}}\left( c\cosh \frac{\tilde{\tau}}c\right) ^2\partial _{\tilde{%
\tau}}+ \\ 
+\frac{f_{(\mu )}\left[ f_{(\nu )}^{\prime }\right] _{,x}}{c_1^2\tanh ^2%
\frac{\tilde{\tau}}c}\partial _x-f_{(\nu )}^{\prime }\left[ f_{(\mu
)}\right] _{,y}\partial _y-\frac{f_{(\nu )}^{\prime }\left[ f_{(\mu
)}\right] _{,z}}{\sinh ^2y}\partial _z
\end{array}
$ \\ \hline
$B_6$ & $1$ & $\pm c_1^2c^2\cosh ^2\frac{\tilde{\tau}}c$ & $\pm c^2\sinh ^2%
\frac{\tilde{\tau}}c$ & {\bf 10} & $
\begin{array}{c}
{\bf X}_{2(\mu +1)+\nu }=f_{(\mu )}\left[ f_{(\nu )}^{\prime }\right] _{,%
\tilde{\tau}}\left( c\sinh \frac{\tilde{\tau}}c\right) ^2\partial _{\tilde{%
\tau}}+ \\ 
+\frac{f_{(\mu )}\left[ f_{(\nu )}^{\prime }\right] _{,x}}{c_1^2\coth ^2%
\frac{\tilde{\tau}}c}\partial _x-f_{(\nu )}^{\prime }\left[ f_{(\mu
)}\right] _{,y}\partial _y-\frac{f_{(\nu )}^{\prime }\left[ f_{(\mu
)}\right] _{,z}}{\sin ^2y}\partial _z
\end{array}
$ \\ \hline
$B_7$ & $0$ & $\pm c_1^2$ & $\pm c_2^2$ & {\bf 10} & $
\begin{array}{c}
{\bf X}_{2(\mu +1)+\nu }=c_2f_{(\mu )}\left[ f_{(\nu )}^{\prime }\right] _{,%
\tilde{\tau}}\partial _{\tilde{\tau}}+\frac{c_2f_{(\mu )}\left[ f_{(\nu
)}^{\prime }\right] _{,x}}{c_1^2}\partial _x- \\ 
-\frac{f_{(\nu )}^{\prime }\left[ f_{(\mu )}\right] _{,y}}{c_2}\partial _y-%
\frac{f_{(\nu )}^{\prime }\left[ f_{(\mu )}\right] _{,z}}{y^2c_2}\partial _z
\\ 
{\bf X}_9=\partial _{\tilde{\tau}} \\ 
{\bf X}_{10}=c_1x\partial _{\tilde{\tau}}-\frac{\tilde{\tau}}{c_1}\partial _x
\end{array}
$ \\ \hline
$B_8$ & $0$ & $\pm c_1^2\tilde{\tau}^2$ & $\pm c_2^2$ & {\bf 10} & $
\begin{array}{c}
{\bf X}_{2(\mu +1)+\nu }=c_2f_{(\mu )}f_{(\nu )}^{\prime }\partial _{\tilde{%
\tau}}+\frac{c_2f_{(\mu )}\left[ f_{(\nu )}^{\prime }\right] _{,x}}{c_1^2%
\tilde{\tau}}\partial _x- \\ 
-\frac{\tilde{\tau}f_{(\nu )}^{\prime }\left[ f_{(\mu )}\right] _{,y}}{c_2}%
\partial _y-\frac{\tilde{\tau}f_{(\nu )}^{\prime }\left[ f_{(\mu )}\right]
_{,z}}{y^2c_2}\partial _z \\ 
{\bf X}_9=\cos c_1x\partial _{\tilde{\tau}}-\frac 1{c_1\tilde{\tau}}\sin
c_1x\partial _x \\ 
{\bf X}_{10}=\sin c_1x\partial _{\tilde{\tau}}+\frac 1{c_1\tilde{\tau}}\cos
c_1x\partial _x
\end{array}
$ \\ \hline
\end{tabular}
\newpage {\bf {\small Table 3}. }{\small Explanations for the quantities }$%
f_{(\mu )},f_{(\nu )}^{\prime }${\small \ appearing in Table 1. Note that }$%
\mu ,\nu =1,2,3$. \vspace{0.1cm}

\begin{tabular}{|l|l|l|l|}
\hline
{\bf Class} & ${\bf k}$ & $f_{(\nu )}^{\prime }$ & $f_{(\mu )}$ \\ \hline
$B_1$ & $1$ & $\left( -\tanh \frac{\tilde{\tau}}c,0,0\right) $ & $\left(
-\cos y,\sin y\cos z,\sin y\sin z\right) $ \\ \hline
$B_2$ & $-1$ & $\left( \coth \frac{\tilde{\tau}}c,0,0\right) $ & $\left(
\cosh y,\sinh y\cos z,\sinh y\sin z\right) $ \\ \hline
$B_3$ & $-1$ & $\left( -\cot \frac{\tilde{\tau}}c,0,0\right) $ & $\left(
\cosh y,\sinh y\cos z,\sinh y\sin z\right) $ \\ \hline
$B_4$ & $1$ & $\left( \tanh \frac{\tilde{\tau}}c\cosh c_1x,\tanh \frac{%
\tilde{\tau}}c\sinh c_1x,0\right) $ & $\left( -\cos y,\sin y\cos z,\sin
y\sin z\right) $ \\ \hline
$B_5$ & $-1$ & $\left( -\coth \frac{\tilde{\tau}}c\cos c_1x,-\coth \frac{%
\tilde{\tau}}c\sin c_1x,0\right) $ & $\left( \cosh y,\sinh y\cos z,\sinh
y\sin z\right) $ \\ \hline
$B_6$ & $-1$ & $\left( -\cot \frac{\tilde{\tau}}c\cosh c_1x,-\cot \frac{%
\tilde{\tau}}c\sinh c_1x,0\right) $ & $\left( \cosh y,\sinh y\cos z,\sinh
y\sin z\right) $ \\ \hline
$B_7$ & $0$ & $-\left( \tilde{\tau},c_1x,0\right) $ & $\left( y\cos z,y\sin
z,0\right) $ \\ \hline
$B_8$ & $0$ & $-\left( \cosh c_1x,\sinh c_1x,0\right) $ & $\left( y\cos
z,y\sin z,0\right) $ \\ \hline
\end{tabular}
\end{center}

\vspace{0.2cm}

\begin{center}
{\bf {\small Table 4}. }{\small Explanations for the quantities }$f_{(\mu
)},f_{(\nu )}^{\prime }${\small \ appeared in Table 2. Note that }$\mu ,\nu
=1,2,3$. \vspace{0.1cm}

\begin{tabular}{|l|l|l|l|}
\hline
{\bf Class} & ${\bf k}$ & $f_{(\nu )}^{\prime }$ & $f_{(\mu )}$ \\ \hline
$B_1$ & $1$ & $\left( -\tan \frac{\tilde{\tau}}c,0,0\right) $ & $\left(
-\cos y,\sin y\cos z,\sin y\sin z\right) $ \\ \hline
$B_2$ & $-1$ & $\left( \tanh \frac{\tilde{\tau}}c,0,0\right) $ & $\left(
\cosh y,\sinh y\cos z,\sinh y\sin z\right) $ \\ \hline
$B_3$ & $1$ & $\left( -\coth \frac{\tilde{\tau}}c,0,0\right) $ & $\left(
\cos y,\sin y\cos z,\sin y\sin z\right) $ \\ \hline
$B_4$ & $1$ & $\left( \tan \frac{\tilde{\tau}}c\cos c_1x,\tan \frac{\tilde{%
\tau}}c\sin c_1x,0\right) $ & $\left( -\cos y,\sin y\cos z,\sin y\sin
z\right) $ \\ \hline
$B_5$ & $-1$ & $\left( -\tanh \frac{\tilde{\tau}}c\cos c_1x,-\tanh \frac{%
\tilde{\tau}}c\sin c_1x,0\right) $ & $\left( \cosh y,\sinh y\cos z,\sinh
y\sin z\right) $ \\ \hline
$B_6$ & $1$ & $\left( -\coth \frac{\tilde{\tau}}c\cosh c_1x,-\coth \frac{%
\tilde{\tau}}c\sinh c_1x,0\right) $ & $\left( \cos y,\sin y\cos z,\sin y\sin
z\right) $ \\ \hline
$B_7$ & $0$ & $-\left( \tilde{\tau},c_1x,0\right) $ & $\left( y\cos z,y\sin
z,0\right) $ \\ \hline
$B_8$ & $0$ & $-\left( \cos c_1x,\sin c_1x,0\right) $ & $\left( y\cos
z,y\sin z,0\right) $ \\ \hline
\end{tabular}
\end{center}

\vspace{0.5cm}

\section{Examples}

\label{SecExamples}

The results of the Tables 1,2 are general and can give the proper RCs and
MCs of any given LRS metric of the type (\ref{sx1.1}). To demonstrate this
and to show their usefulness we consider the following examples.

\subsection{The homogeneous and self-similar LRS metrics (1.1)}

One application of the results of Table 1 is the determination of all LRS
metrics of type (\ref{sx1.1}) which are homogeneous, that is they accept
additional KVs. These spacetimes have been determined previously in \cite
{Apostol-Tsamp5} using the reduction of a CKV to a KV by vanishing the
conformal factor. However here the procedure is different, that is, one uses
the component $g_0 $ to compute $\bar{\tau}$ in terms of $t$ and then
replace the result in the expressions for $K_1,K_2$ and ${\bf X}$. The
results of the calculations are given in Table 5 (see also Table 3 p.3780 
\cite{Apostol-Tsamp5}), which we include for convenience and completeness of
the present study. For the same reason we give in Table 6 the self similar
metrics of type (\ref{sx1.1}), which have been determined by Sintes \cite
{Sintes1}. These spacetimes do not admit proper RCs and MCs. \clearpage

\begin{center}
{\small Table 5. This Table contains all homogeneous LRS spacetimes with
metric (1.1). The indices }$\alpha ,\mu =2,3${\small \ and the constants }$%
a,c,\varepsilon _1${\small \ satisfy the constraints }$ac\neq 0${\small \
and }$\varepsilon _1=\pm 1${\small .}

\begin{tabular}{|l|l|l|l|l|l|}
\hline
{\bf Case} & $k$ & $A(t)$ & $B(t)$ & {\bf KVs} & {\bf Type of the metric} \\ 
\hline
A$_1$ & $0$ & $ce^{-t/c\alpha _1}$ & $ce^{-t/c}$ & ${\bf \xi }$ & LRS \\ 
\hline
A$_2$ & $\pm 1$ & $c$ & $c$ & ${\bf \xi }_\mu $ & 1+1+2 \\ \hline
A$_3$ & $0,\pm 1$ & $ce^{\varepsilon _1t/ac}$ & $c$ & ${\bf \xi }_\mu $ & 2+2
\\ \hline
A$_4$ & $0,\pm 1$ & $c\cosh \frac t{ca}$ & $c$ & ${\bf \xi }_\mu $ & 2+2 \\ 
\hline
A$_5$ & $0,\pm 1$ & $c\sinh \frac t{ca}$ & $c$ & ${\bf \xi }_\mu $ & 2+2 \\ 
\hline
A$_6$ & $0,\pm 1$ & $c\cos \frac t{ca}$ & $c$ & ${\bf \xi }_\mu $ & 2+2 \\ 
\hline
A$_7$ & $\pm 1$ & $ct$ & $c$ & ${\bf \xi }_\mu $ & 1+1+2 \\ \hline
B$_1$ & $1$ & $c\cot \tau $ & $\frac c{\sin \tau }$ & ${\bf X}_{3(\alpha
+1)+\mu }$ & Constant Curvature (Type a) \\ \hline
B$_2$ & $-1$ & $c\coth \tau $ & $\frac c{\sinh \tau }$ & ${\bf X}_{3(\alpha
+1)+\mu }$ & Constant Curvature (Type a) \\ \hline
B$_3$ & $-1$ & $c\tanh \tau $ & $\frac c{\cosh \tau }$ & ${\bf X}_{3(\alpha
+1)+\mu }$ & Constant Curvature (Type a) \\ \hline
B$_1$ & $1$ & $c$ & $\frac c{\cos \tau }$ & ${\bf X}_{6+\mu }$ & 1+3 (Type b)
\\ \hline
B$_3$ & $-1$ & $c$ & $\frac c{\cosh \tau }$ & ${\bf X}_{6+\mu }$ & 1+3 (Type
b) \\ \hline
B$_4$ & $-1$ & $c$ & $\frac c{\sinh \tau }$ & ${\bf X}_{6+\mu }$ & 1+3 (Type
b) \\ \hline
B$_1$ & $1$ & $c\cos \frac tc$ & $c$ & ${\bf X}_\mu $ & 2+2 (Type c) \\ 
\hline
B$_2$ & $-1$ & $ce^{\varepsilon _1t/c}$ & $c$ & ${\bf X}_\mu $ & 2+2 (Type c)
\\ \hline
B$_3$ & $-1$ & $c\cosh \frac tc$ & $c$ & ${\bf X}_\mu $ & 2+2 (Type c) \\ 
\hline
B$_4$ & $-1$ & $c\sinh \frac tc$ & $c$ & ${\bf X}_\mu $ & 2+2 (Type c) \\ 
\hline
\end{tabular}
\bigskip

{\small Table 6. LRS spacetimes (\ref{sx1.1}) with transitive homothety
group }$H_5${\small \ and }$\varepsilon _1=\pm 1${\small .}

\begin{tabular}{|l|l|l|l|l|l|}
\hline
{\bf Case} & $k$ & $A(t)$ & $B(t)$ & {\bf HKVs} & {\bf Conformal Factor} \\ 
\hline
A$_1$ & $0$ & $t^{\frac{b-1}b}$ & $c_1t^{\frac{b-\alpha _1}b}$ & ${\bf \xi }%
=bt\partial _t+x\partial _x+\alpha _1y\partial _y$ & $b$ \\ \hline
A$_2$ & $\pm 1$ & $\alpha _1t$ & $\alpha _1t$ & ${\bf \xi }=\alpha
_1t\partial _t$ & $\alpha _1$ \\ \hline
A$_3$ & $0,\pm 1$ & $(\alpha _1t)^{(1+\frac{\varepsilon _1}{c_1a})}$ & $%
\alpha _1t$ & ${\bf \xi }=-\varepsilon _1a\alpha _1t\partial _t+x\partial _x$
& $-\varepsilon _1a\alpha _1$ \\ \hline
\end{tabular}
\end{center}

\subsection{The Datta solution}

The Einstein-Maxwell spacetimes admitting a $G_3I$ on spacelike 3D
hypersurfaces have been given explicitly by Datta \cite{Datta}. Physically
these solutions can be used to model cosmologies with a cosmic magnetic
field. Geometrically they contain - among other - plane symmetric LRS models
with symmetry group $G_4\supset G_3I$) and metric: 
\begin{equation}
ds^2=-\frac{dt^2}{bt^{-1}-at^{-2}}+\left( bt^{-1}-at^{-2}\right)
dx^2+t^2\left( dy^2+y^2dz^2\right)  \label{ex6}
\end{equation}
Using the results of Tables 1,2 we shall determine the proper RCs and the
proper MCs of these LRS solutions.

We compute the Ricci tensor (which is identical for these metrics with the
Einstein tensor): 
\begin{equation}
R_{ab}=\frac a{t^2\left( bt-a\right) }dt^2+\frac{a\left( a-bt\right) }{t^6}%
dx^2+\frac a{t^2}\left( dy^2+y^2dz^2\right) .  \label{sx7}
\end{equation}
An examination of Tables 1,2 shows that the only possible case for the
spacetime (\ref{ex6}) to admit a proper RC/MC is case A$_1$ with
collineation vector ${\bf X}=\alpha _1c\partial _{\tilde{\tau}}+x\partial
_x+\alpha _1y\partial _y$. Taking the Lie derivative of $R_{ab}$ (\ref{sx7})
w.r.t. ${\bf X}$ and setting equal to zero we find $b=0,c=1,\alpha _1=1/3$.
But for these values ${\bf X}$ becomes a HVF with homothetic factor $\psi
=2/3$. We conclude that the Einstein-Maxwell plane symmetric models (\ref{ex6}%
) do not admit proper RCs and MCs.

\subsection{The stiff perfect fluid LRS solution}

\label{SubSecExamples2}

Another example of LRS spacetime (\ref{sx1.1}) is the stiff fluid model ($%
\gamma =2$) \cite{Kramer} with metric: 
\begin{equation}
ds^2=-dt^2+t^{2/(1+2\lambda )}dx^2+t^{2\lambda /(1+2\lambda )}\left(
dy^2+y^2dz^2\right) .  \label{sx8}
\end{equation}
The Einstein tensor is computed to be (the Ricci tensor is degenerate!): 
\begin{equation}
{\bf G}=\frac{\lambda (\lambda +2)}{t^2(2\lambda +1)^2}dt^2+t^{-4\lambda
/(2\lambda +1)}\frac{\lambda (\lambda +2)}{(2\lambda +1)^2}%
dx^2+t^{-2(\lambda +1)/(2\lambda +1)}\frac{\lambda (\lambda +2)}{(2\lambda
+1)^2}\left( dy^2+y^2dz^2\right) .  \label{sx9}
\end{equation}
From Tables 1,2 we find that the only possible\footnote{%
The others are easily excluded. E.g. A$_4$ implies that $G_2=t^{-2(\lambda
+1)/(2\lambda +1)}\frac{\lambda (\lambda +2)}{(2\lambda +1)^2}=const$, which
in turn gives $\lambda +1=0$ which is not acceptable because the solution (%
\ref{sx8}) is defined for $\lambda >0$ or $\lambda <-2$} case for a proper
MC to be admitted is case A$_1$, the collineation vector being ${\bf X}%
=\alpha _1c\partial _{\tilde{\tau}}+x\partial _x+\alpha _1y\partial _y$. By
demanding ${\cal L}_{{\bf X}}G_{ab}=0$ (and recalling that $d\tilde{\tau}%
=\left| G_0\right| ^{1/2}dt$) we find $\alpha _1=\frac{\lambda +1}{2\lambda }
$ and $c=\frac{\sqrt{\lambda (\lambda +2)}}\lambda $. However for these
values of the parameters it is easy to show that the collineation ${\bf X}$
reduces to a HVF with homothetic factor $\psi =\frac{2\lambda +1}\lambda $.
We conclude that the $\gamma =2$ (plane symmetric) model (\ref{sx8}) does
not admit proper MCs.

\subsection{RCs and MCs of static spherically symmetric spacetimes}

\label{SubSecStaticSTs}

The static spherically symmetric spacetimes are a special and interesting
class of LRS spacetimes of type (\ref{sx1.1}). The problem of finding all
static spherically symmetric spacetimes admitting RCs has been considered
many times in the literature \cite
{Bokhari-Qadir,Amir-Bokhari-Qadir,Farid-Qadir-Ziad,Bertolotti-Contreras-Nunez-Percoco-Carot,Ziad}%
. The results of all these works follow immediately from Tables 1,2 by
considering the $k=1,\epsilon =1$ cases and interchanging $x\leftrightarrow
t $. Table 7 shows the correspondence between the cases resulting from
Tables 1,2 with the most recent and complete work on this topic \cite{Ziad}.

Concerning the determination of the static spherically symmetric spacetimes
which admit MCs these follow immediately from Tables 1,2 without any further
calculations. These results are new. Indeed the results in the current
literature \cite{Sharif1} concern very special cases of LRS spacetimes (
(anti) de Sitter spacetime,Bertotti-Robinson spacetime,
anti-Bertotti-Robinson spacetime, (anti) Einstein spacetime, Schwarzschild
spacetime and Reissner-Nordstrom spacetime) in which either there no proper
MCs or their Lie algebra is infinite dimensional (degenerate case).

\begin{center}
{\small Table 7. Comparison of the results found in the present paper
concerning RCs in static spherically symmetric spacetimes with known results
from the literature. }\\[0pt]

\begin{tabular}{|l|l|l|}
\hline
{\bf Case of the present paper} & {\bf Reference} & {\bf Result} \\ \hline
A$_2$,A$_3$,A$_4$,A$_5$,A$_6$,A$_7$ & \cite{Ziad} & Theorem 3 \\ \hline
B$_1$,B$_3$ & \cite{Ziad} & Theorem 5 \\ \hline
B$_4$,B$_6$ & \cite{Ziad} & Theorem 6 \\ \hline
\end{tabular}
\end{center}

\section{Ricci and Matter Collineations of the LRS metrics (1.2)}

Concerning the metrics (\ref{sx1.2}) working as previously we compute the
Ricci tensor and the Einstein tensor. We find: 
\begin{eqnarray}
R_{00} &=&R_0=-\frac{\ddot{A}B+2\ddot{B}A}{AB}  \nonumber \\
R_{11} &=&R_1=\frac{A^4}{2B^4}+\frac{2\dot{B}\dot{A}A}B+\ddot{A}A  \nonumber
\\
R_{13} &=&R_1\Lambda  \label{sx3.01} \\
R_{22} &=&R_2=-\frac{A^2(\Lambda _{,y})^2}{2B^2\Sigma ^2}+\frac{\dot{B}\dot{A%
}B}A+\ddot{B}B+\dot{B}^2+k  \nonumber \\
R_{33} &=&\left( R_1\Lambda ^2+R_2\Sigma ^2\right)  \nonumber
\end{eqnarray}
\begin{eqnarray}
G_{00} &=&G_0=-\frac{A^2(\Lambda _{,y})^2}{4B^4\Sigma ^2}+\frac{2\dot{B}\dot{%
A}}{AB}+\frac{\dot{B}^2}{B^2}+\frac k{B^2}  \nonumber \\
G_{11} &=&G_1=\frac{3A^4(\Lambda _{,y})^2}{4B^4\Sigma ^2}-\frac{2\ddot{B}A^2}%
B-\frac{\dot{B}^2A^2}{B^2}-k\frac{A^2}{B^2}  \nonumber \\
G_{13} &=&G_1\Lambda  \label{sx3.02} \\
G_{22} &=&G_2=-\frac{A^2(\Lambda _{,y})^2}{4B^2\Sigma ^2}-\frac{\ddot{A}B^2}%
A-\frac{\dot{B}\dot{A}B}A-\ddot{B}B  \nonumber \\
G_{33} &=&\left( G_1\Lambda ^2+G_2\Sigma ^2\right)  \nonumber
\end{eqnarray}
We observe that the form of $R_{ab},G_{ab}$ is similar to that of the metric 
$g_{ab}$, therefore it is possible to consider the ``generic'' metric: 
\begin{equation}
ds^2=\varepsilon _1d\tilde{\tau}^2+\varepsilon _2\left| K_1\right| \left[
dx+\Lambda (y,k)dz\right] ^2+\varepsilon _3\left| K_2\right| \left[
dy^2+\Sigma ^2(y,k)dz^2\right]  \label{sx3.1}
\end{equation}
where $K_\alpha =\{g_\alpha ,R_\alpha ,G_\alpha \}$ and $\varepsilon
_1,\varepsilon _2,\varepsilon _3=\pm 1$ are the signs of $K_0,K_1,K_2$
components respectively.

As in the case of LRS metrics (\ref{sx1.1}) we have to consider two cases,
i.e. the non-conformally flat and the conformally flat ''generic'' metrics (%
\ref{sx3.1}).

Concerning the non-conformally flat we have the following result (see the
similar result in \cite{Apostol-Tsamp5}; the proof is given in Appendix 1):

{\it The LRS spacetimes (\ref{sx1.2}) admit at most one proper RC or MC
given by: } 
\begin{equation}
{\bf X}=\partial _{\tilde{\tau}}+2ax\partial _x+ay\partial _y  \label{sx3.2}
\end{equation}
{\it in which case the components of the Ricci and the Einstein tensor
satisfy the constraints: } 
\begin{equation}
K_1=\pm \left( c_1e^{-2a\tilde{\tau}}\right) ^2\qquad K_2=\pm \left( c_1e^{-a%
\tilde{\tau}}\right) ^2  \label{sx3.3}
\end{equation}
{\it where }$a=0${\it \ when }$k\neq 0${\it \ \ and }$c_1,c_2${\it \ \ are
non-vanishing constants. Furthermore }$c_1\neq 1${\it \ \ for }$a=0${\it \ \
in order to avoid the conformally flat case. }

In the conformally flat cases we show in Appendix 2 that the functions $%
K_1,K_2 $\ satisfy $K_1=kK_2$ ($k\neq 0$) and the ``generic'' metric becomes
conformally related to a 1+3 decomposable metric. The KVs are then
determined easily (see for example the method developed in \cite
{Tsa-Nik-Apos}, or \cite{Capocci-Hall1}). The calculations are standard and
there is no need to be referred explicitly. The result is that in this case
(i.e $K_1=kK_2$ ($k\neq 0$)) there are two proper RCs and two proper MCs
given by the following vectors: 
\begin{equation}
{\bf X}_1=-k\frac{\Lambda ^{\prime }}\Lambda \cos x\partial _x+\sin
x\partial _y-\Lambda ^{-1}\cos x\partial _z  \label{sx3.4}
\end{equation}
\begin{equation}
{\bf X}_2=k\frac{\Lambda ^{\prime }}\Lambda \sin x\partial _x+\cos x\partial
_y+\Lambda ^{-1}\sin x\partial _z.  \label{sx3.5}
\end{equation}

In case the spacetime admits extra RCs and MCs the components $K_1$ (or $K_2$%
) satisfy additional restrictions. The analysis shows that in this case the
component $K_1$ takes one of the following forms: 
\begin{equation}
K_1=\sinh ^2\frac{\tilde{\tau}}2,\qquad K_1=\cosh ^2\frac{\tilde{\tau}}%
2,\qquad K_1=\sin ^2\frac{\tilde{\tau}}2  \label{sx3.9}
\end{equation}
and, furthermore, that there are {\em four} extra RCs or MCs (proper or not)
as follows: 
\begin{equation}
{\bf X}_{(n)}=2\lambda _{(n)}\partial _{\tilde{\tau}}+4k\cdot \left( \ln
K_1\right) _{,\tilde{\tau}}\left[ \lambda _{(n)}\right] _{,\alpha
}F_{(n)}^{\alpha \beta }\partial _\beta  \label{sx3.6}
\end{equation}
where the quantities $\lambda _{(n)}$, $F^{\alpha \beta }$ are given by: 
\begin{equation}
\begin{tabular}{l}
$\lambda _1=\left[ \Lambda (y,k)+1\right] ^{1/2}\sin (\frac x2+\frac z2)$ \\ 
$\lambda _2=\left[ \Lambda (y,k)+1\right] ^{1/2}\cos (\frac x2+\frac z2)$ \\ 
$\lambda _3=\left[ 1-\Lambda (y,k)\right] ^{1/2}\sin (\frac x2-\frac z2)$ \\ 
$\lambda _4=\left[ 1-\Lambda (y,k)\right] ^{1/2}\cos (\frac x2-\frac z2)$%
\end{tabular}
\label{sx3.7}
\end{equation}
\begin{equation}
F_{(1,2)}^{\alpha \beta }=diag\left( \frac 1{\Lambda (y,k)+1},k,\frac
1{\Lambda (y,k)+1}\right)  \label{sx3.8}
\end{equation}
\begin{equation}
F_{(3,4)}^{\alpha \beta }=diag\left( \frac 1{1-\Lambda (y,k)},k,\frac
1{1-\Lambda (y,k)}\right) .  \label{sx3.8b}
\end{equation}

\section{Ricci and Matter Collineations of the LRS metrics (1.3)}

In this section we continue with the remaining LRS metrics (\ref{sx1.3}). We
compute the Ricci and the Einstein tensor and we find: 
\begin{eqnarray}
R_{00} &=&-\frac{2\ddot{B}A+\ddot{A}B}{AB}=R_0  \nonumber \\
R_{01} &=&2\left( \frac{\dot{A}}A-\frac{\dot{B}}B\right)  \label{sx4.001} \\
R_{11} &=&\frac AB\left( \ddot{A}B+2\dot{B}\dot{A}\right) -2=R_1  \nonumber
\\
R_{22} &=&R_{33}=e^{2x}\left( \frac{\dot{B}\dot{A}B}A-2\frac{B^2}{A^2}+\ddot{%
B}B+\dot{B}^2\right) =e^{2x}R_2=e^{2x}R_3  \nonumber
\end{eqnarray}
\begin{eqnarray}
G_{00} &=&\frac{\dot{B}^2A^2+2\dot{B}\dot{A}AB-3B^2}{A^2B^2}=G_0  \nonumber
\\
G_{01} &=&2\left( \frac{\dot{A}}A-\frac{\dot{B}}B\right)  \label{sx4.002} \\
G_{11} &=&\frac{B^2-A^2\left( 2\ddot{B}B+\dot{B}^2\right) }{B^2}=G_1 
\nonumber \\
G_{22} &=&G_{33}=-e^{2x}\frac B{A^2}\left( \ddot{B}A^2+A\ddot{A}B+A\dot{B}%
\dot{A}-B\right) =e^{2x}G_2=e^{2x}G_3  \nonumber
\end{eqnarray}
We observe that the tensors $R_{ab},G_{ab}$ are not (in general) diagonal
therefore we cannot consider (in general) a generic metric as we did for the
previous cases and we have to work differently.

The form of the metric indicates that we must consider two cases, that is, $%
R_{ab},G_{ab}$ \ diagonal and not diagonal. \newline
\underline{Case\ I:\ \ $G_{ab},R_{ab}$\ diagonal}

The requirement $G_{ab},R_{ab}$\ diagonal gives the condition $A(t)=cB(t)$
from which it follows easily that the metric (\ref{sx1.3}) is conformally
flat. Although we cannot consider a generic metric element for all three
metrics $ds_g^2,ds_R^2,ds_T^2$ it is still possible to consider one such for
the two metric elements\footnote{%
Assuming that both $R_{ab}$ and $G_{ab}$ are non-degenerate.} $ds_R^2$ and $%
ds_T^2$ as follows: 
\begin{equation}
ds_{R-T}^2=K_0dt^2+K_1dx^2+K_2e^{2x}\left( dy^2+dz^2\right)  \label{sx4.003}
\end{equation}
where $K_a=\{R_a,G_a\}$ ($R_a,T_a$) being defined in (\ref{sx4.001}) and (%
\ref{sx4.002}). We note that $ds_{R-T}^2$ need not be conformally flat. By
means of the transformation: 
\begin{equation}
d\tilde{\tau}=\left| K_0\right| ^{1/2}dt  \label{sx4.004}
\end{equation}
the generic metric takes the form: 
\begin{equation}
ds^2=\varepsilon _1(K_0)d\tilde{\tau}^2+K_1dx^2+K_2e^{2x}\left(
dy^2+dz^2\right) .  \label{sx4.005}
\end{equation}

For the determination of the KVs of the generic metric $ds_{R-T}^2$ we have
to consider, as previously, two cases, that is, the non-conformally flat and
the conformally flat case. \newline
\underline{Case IA. $ds_{R-T}^2$ non-conformally flat }

For the Lorentzian signature $\varepsilon =-1$ it has been shown \cite
{Apostol-Tsamp5} that there exist two KVs\footnote{%
We note that in \cite{Apostol-Tsamp5} a second KV has been omitted.}. For
the remaining Euclidean signature we determine the KVs in the standard way
in which there exists at most one KV for the generic metric. The results for
all cases are collected in Table 8.

\begin{center}
{\bf {\small Table 8}.}{\small \ Proper RCs and Proper MCs for the case IA} 
\\[0pt]
\begin{tabular}{|l|l|l|l|l|}
\hline
Class & $K_{1}$ & $K_{2}$ & sign$K_{0}K_{1}$ & ${\bf X}$ \\ \hline
$A_{1}$ & $\sinh ^{2}\tilde{\tau}$ & $\sinh ^{-2}\tilde{\tau}$ & $-1$ & $%
e^{-x}\left( \partial _{\tilde{\tau}}+\coth \tilde{\tau}\partial _{x}\right) 
$ \\ \hline
$A_{2}$ & $\cos ^{2}\tilde{\tau}$ & $\cos ^{-2}\tilde{\tau}$ & $-1$ & $%
e^{-x}\left( \partial _{\tilde{\tau}}-\tan \tilde{\tau}\partial _{x}\right) $
\\ \hline
$A_{3}$ & $\cosh ^{2}\tilde{\tau}$ & $\cosh ^{-2}\tilde{\tau}$ & $\; 1$ & $%
e^{-x}\left( \partial _{\tilde{\tau}}+\tanh \tilde{\tau}\partial _{x}\right) 
$ \\ \hline
\end{tabular}
\end{center}

\underline{Case IB. $ds_{R-T}^2$ conformally flat}

The condition for $ds_{R-T}^2$ to be conformally flat is: 
\begin{equation}
\left[ \frac{\left( K_2\right) _{,\tilde{\tau}}}{K_2}\right] _{,\tilde{\tau}}%
\frac{K_1}{K_2}=\left[ \frac{\left( K_1\right) _{,\tilde{\tau}}}{K_2}\right]
_{,\tilde{\tau}}.  \label{sx4.006}
\end{equation}
This condition leads to the following solutions: 
\begin{equation}
K_1=cK_2\qquad or\qquad K_1=ce^{\int \frac{d\tilde{\tau}}{cK_1}}K_2.
\label{sx4.007}
\end{equation}
It is found (see also the corresponding result in \cite{Apostol-Tsamp5})
that in both cases there exists one proper MC or one proper RC given by the
following vector: 
\begin{eqnarray}
\varepsilon _1 &=&1,\qquad {\bf X}=K_1\partial _{\tilde{\tau}}
\label{sx4.008a} \\
\varepsilon _1 &=&-1,\qquad {\bf X}=cK_1\partial _{\tilde{\tau}}+\partial _x.
\label{sx4.008b}
\end{eqnarray}
\underline{Case II: Non Diagonal $G_{ab},R_{ab}$}

In this case there are no shortcuts to apply and we must solve directly the
collineation equations. We introduce the functions $v(t),w(t)$ as follows: 
\begin{equation}
B(t)=e^{v(t)}\qquad ,\qquad \frac{A(t)}{B(t)}=e^{w(t)}  \label{sx4.01}
\end{equation}
where we assume that $w\cdot w_{,t}\ne 0$ in order to avoid the case $%
G_{ab},R_{ab}$ diagonal. Next we introduce the new coordinate $\tilde{\tau}$
by the relation: 
\begin{equation}
dt=\frac{d\tilde{\tau}}{2w_{,t}}\Leftrightarrow \tilde{\tau}=w(t).
\label{sx4.02}
\end{equation}
Using equations (\ref{sx4.01}) and (\ref{sx4.02}) the metric (\ref{sx1.3})
becomes: 
\begin{equation}
ds^2=-\frac{d\tilde{\tau}^2}{(2w_{,t})^2}+e^{2\left[ \tilde{\tau}%
+v(t)\right] }dx^2+e^{2\left[ x+v(t)\right] }\left( dy^2+dz^2\right) .
\label{sx4.1}
\end{equation}
The reason for introducing the new coordinate $\tilde{\tau}$ is that in the
coordinate system $\left\{ \tilde{\tau},x,y,z\right\} $ the non-diagonal
component of $R_{ab}$ and $G_{ab}$ is constant. Indeed the non-vanishing
components of the Ricci and Einstein tensor in these coordinates are: 
\begin{eqnarray}
R_{00} &=&-\left[ 3\left( \ddot{v}+\dot{v}+\frac{\ddot{w}\dot{v}}{\dot{w}}%
\right) +2\dot{v}+\frac{\ddot{w}}{\dot{w}}-1\right] =R_0  \nonumber \\
R_{01} &=&2  \label{sx4.2} \\
R_{11} &=&4e^{2\left[ \tilde{\tau}+v(t)\right] }\left( \ddot{v}\dot{w}^2+3%
\dot{w}^2\dot{v}^2+\dot{v}\dot{w}\ddot{w}+4\dot{w}^2+\dot{w}\ddot{w}+\dot{w}%
^2\right) -2=R_1  \nonumber \\
R_{22} &=&R_{33}=4e^{2\left[ x+v(t)\right] }\left( \ddot{v}\dot{w}^2+3\dot{w}%
^2\dot{v}^2+\dot{v}\dot{w}\ddot{w}+4\dot{w}^2\right) -2e^{2(x-\tilde{\tau}%
)}=e^{2x}R_2  \nonumber
\end{eqnarray}
\begin{eqnarray}
G_{00} &=&-\frac{3e^{-2\left[ \tilde{\tau}+v(t)\right] }}{\dot{w}^2}+3\dot{v}%
^2+2\dot{v}=G_0  \nonumber \\
G_{01} &=&2  \label{sx4.3} \\
G_{11} &=&-4e^{2\left[ \tilde{\tau}+v(t)\right] }\left( 2\ddot{v}\dot{w}^2+3%
\dot{w}^2\dot{v}^2+2\dot{v}\dot{w}\ddot{w}\right) +1=G_1  \nonumber \\
G_{22} &=&G_{33}=-4e^{2\left[ x+v(t)\right] }\left( \ddot{v}\dot{w}^2+3\dot{w%
}^2\dot{v}^2+2\dot{v}\dot{w}\ddot{w}+3\dot{w}^2\dot{v}+\dot{w}\ddot{w}+\dot{w%
}^2\right) +1=e^{2x}G_2  \nonumber
\end{eqnarray}
where a dot denotes differentiation w.r.t. the new coordinate $\tilde{\tau}$.

As in the last case the Ricci and the Einstein tensor ({\em but not the
metric $g_{ab}$}!) follow as particular cases of a new ``generic'' metric
element: 
\begin{equation}
ds_{R-T}^{\prime 2}=K_0d\tilde{\tau}^2+4dxd\tilde{\tau}+K_1dx^2+K_2e^{2x}%
\left( dy^2+dz^2\right) .  \label{sx4.3a}
\end{equation}
whose KVs will produce all proper RCs and MCs, if there exist. In order to
solve Killing's equations for the metric $ds_{R-T}^{\prime 2}$ we consider
again subcases according to whether the metric $ds_{R-T}^{\prime 2}$ is
conformally flat or not. \newline
\underline{Case IIA: $ds_{R-T}^{\prime 2}$ non-conformally flat}

In this case the metric element $ds_{R-T}^{\prime 2}$ is written: 
\begin{equation}
ds_{R-T}^{\prime 2}=K_2e^{2x}\left[ K_2^{-1}e^{-2x}\left( K_0d\tilde{\tau}%
^2+4dxd\tilde{\tau}+K_1dx^2\right) +\left( dy^2+dz^2\right) \right]
\label{sx4.3b}
\end{equation}
that is, it becomes conformal to a 2+2 decomposable metric. It is well known 
\cite{Apostol-Tsamp5,Capocci-Hall1,Carot-Tupper} that the KVs of a 2+2
decomposable metric are identical with the KVs of the constituent 2-metrics.
Therefore the only possible KVs can come from the KVs of the 2-space $\tilde{%
\tau},x$. It is also well known that if a 2-metric admits 2 KVs then it
admits 3 and it is a metric of constant curvature. However if the 2-space is
of constant curvature then it must be flat (because the scalar curvature
contains the factor $e^{2x}$). If this is the case the 4-metric is
conformally flat, which contradicts our assumption. We conclude that there
exists at most one KV (in addition to the four KVs given in (\ref{sx2.1}),
which give rise to trivial RCs and MCs and do not interest us), which must
be of the form: 
\begin{equation}
{\bf X}=X^0(\tilde{\tau},x)\partial _{\tilde{\tau}}+X^1(\tilde{\tau}%
,x)\partial _x  \label{sx4.4}
\end{equation}
where $X^0(\tilde{\tau},x),X^1(\tilde{\tau},x)$ are smooth functions of
their arguments and $X^1\ne 0$ because it leads to a degenerate $K_{ab}$.

To determine the functions $X^0,X^1$ we use Killing's equations. They read: 
\begin{eqnarray}
C_{00} &:&\qquad \dot{K}_0X^0+2K_0\dot{X}^0+4\dot{X}^1=0  \nonumber \\
C_{01} &:&\qquad K_0\left( X^0\right) _{,x}+2\dot{X}^0+2(X^1)_{,x}+K_1\dot{X}%
^1=0  \label{sx4.5} \\
C_{11} &:&\qquad \dot{K}_1X^0+4\left( X^0\right) _{,x}+2K_1(X^1)_{,x}=0 
\nonumber \\
C_{22} &:&\qquad \dot{K}_2X^0+2K_2X^1=0  \nonumber
\end{eqnarray}

We consider two subcases depending on the vanishing of $\left( X^0\right)
_{,x}, \left( X^1\right) _{,x}$.

\underline{Subcase IIA,1: $\left( X^0\right) _{,x}=\left( X^1\right) _{,x}=0$%
}

Equation C$_{11}$ gives $\dot{K}_1=0\Rightarrow K_1=$const. Then from C$%
_{01} $ we obtain: 
\begin{equation}
X^0=\frac{c_1-X^1K_1}2  \label{sx4.6}
\end{equation}
where $c_1$ is a constant of integration.

Replacing $X^0$ back to C$_{22}$ and demanding $\frac{\dot{K}_2}{K_2}\frac{%
K_1}2-2\neq 0$ (in order to avoid the the conformal flatness of $ds_{R-G}^2$%
) we find: 
\begin{equation}
X^1=\frac{\dot{K}_2}{K_2}\frac{c_1}2\left( \frac{\dot{K}_2}{K_2}\frac{K_1}%
2-2\right) ^{-1}.  \label{sx4.7}
\end{equation}
Finally using C$_{00}$ and (\ref{sx4.6}) we obtain ${\bf {X}}$ as well as
the constraints on the metric components for ${\bf {X}}$ to be admitted. 
\footnote{%
We note that $K_0\neq 4/K_1 \Leftrightarrow \det K_{ab}\neq 0 $.} 
\begin{equation}
{\bf X}=\frac{c_2}{\left| K_0-4/K_1\right| ^{1/2}}\partial _{\tilde{\tau}}+%
\frac{\dot{K}_2}{K_2}\frac{c_1}2\left( \frac{\dot{K}_2}{K_2}\frac{K_1}%
2-2\right) ^{-1}\partial _x  \label{sx4.8}
\end{equation}
\begin{equation}
K_1=const.,\qquad \frac{c_2}{\left| K_0-4/K_1\right| ^{1/2}}=\frac{c_1}2-%
\frac{\dot{K}_2}{K_2}\frac{c_1}4\left( \frac{\dot{K}_2}{K_2}\frac{K_1}%
2-2\right) ^{-1}.  \label{sx4.9}
\end{equation}
where $c_2$ is a constant of integration.

\underline{Subcase IIA,2: $\left( X^0\right) _{,x},\left( X^1\right)
_{,x}\neq 0$}

In this case from C$_{22}$ we obtain $K_2=D_2e^{c_2\tilde{\tau}}$ and $X^1=-%
\frac{c_2}2X^0$. Then C$_{11}$ gives $K_1=\frac{4\pm e^{-c_1c_2\tilde{\tau}}%
}{c_2}$ (note that $K_1\neq 4/c_2$ in order to avoid the conformally flat
case) and $X^0=e^{-c_1x}+f(\tilde{\tau})$ where $f(\tilde{\tau})$ is a
smooth function. From C$_{01}$ we compute the function $f(\tilde{\tau})=\ln
\frac 1{D_0\left| K_0-c_2\right| ^{1/2}}$ in terms of $K_0$. It remains
equation C$_{00}$ which gives $K_0=c_2-\frac 1{\frac 4{c_2}e^{c_1c_2\tilde{%
\tau}}+D_3}$ where $D_0,D_3,c_1,c_2$ are constants of integration.

Therefore in this case we have the collineation: 
\begin{equation}
{\bf X}=\left( e^{-c_1x}+\ln \frac 1{D_0\left| K_0(\tilde{\tau})-c_2\right|
^{1/2}}\right) \left( \partial _{\tilde{\tau}}-\frac{c_2}2\partial _x\right)
\label{sx4.10}
\end{equation}
under the conditions: 
\begin{equation}
K_0=c_2-\frac 1{\frac 4{c_2}e^{c_1c_2\tilde{\tau}}+D_3}\qquad ,K_1=\frac{%
4\pm e^{-c_1c_2\tilde{\tau}}}{c_2}\qquad ,K_2=D_2e^{c_2\tilde{\tau}}.
\label{sx4.11}
\end{equation}
\newline

\underline{Case IIB: $ds_{R-G}^{\prime 2}$ conformally flat }

In order the metric $ds_{R-T}^{\prime 2}$ to be conformally flat the
2-dimensional metric: 
\begin{equation}
ds_2^2=e^{-2x}K_1^{-1}\left[ K_0dt^2+4dtdx+K_2dx^2\right]  \label{sx4.12}
\end{equation}
where $K_0(t),K_1(t),K_2(t)$ are smooth functions of $t$ must be flat. This
implies the condition: 
\begin{eqnarray}
&&K_2K_1\dot{K}_0\left( K_2\dot{K}_1-K_1\dot{K}_2+4K_2\right) +2\ddot{K}%
_1K_2^2\left( 4-K_1K_0\right)  \nonumber \\
&&+K_0K_2^2\dot{K}_1^2+K_2\dot{K}_1\left[ \dot{K}_2\left( K_1K_0-8\right)
+4K_2K_0\right] +2K_1\left( K_1K_0-4\right) \left( K_2\ddot{K}_2-\dot{K}%
_2^2\right) =0  \label{sx4.13}
\end{eqnarray}
Furthermore the 2-metric must admit 3 independent KVs of the form: 
\begin{equation}
{\bf X}=X^0(t,x)\partial _t+X^1(t,x)\partial _x  \label{sx4.14}
\end{equation}
where $X^0(t,x),X^1(t,x)$ are at least $C^1$ functions. To determine the the
functions $X^0(t,x),X^1(t,x)$ we use Killing's equations ${\cal L}_{{\bf X}%
}ds_2^2=0$. We compute:

\begin{equation}
C_{00}:-\left( \dot{K}_0K_2-K_0\dot{K}_2\right)
X^0+2K_2K_0X_{,t}^0+4K_2X_{,t}^1-2K_2K_0X^1=0  \label{sx4.15}
\end{equation}
\begin{equation}
C_{01}:K_2\left[ K_0X_{,x}^0+2X_{,t}^0+2X_{,x}^1+K_1X_{,t}^1-4X^1\right] -%
\dot{K}_2X^0=0  \label{sx4.16}
\end{equation}
\begin{equation}
C_{11}:X^0\left( K_2\dot{K}_1-K_1\dot{K}_2\right)
+4K_2X_{,x}^0+4K_2K_1X_{,x}^1-2K_2K_1X^1=0  \label{sx4.17}
\end{equation}
where a dot denotes differentiation w.r.t. $t$. Unfortunately we have not
been able to solve the system of pdf's (\ref{sx4.13}), (\ref{sx4.15})-(\ref
{sx4.17}).

\section{Conclusions}

Following the method of the generic metric, proposed in the introduction, we
have been able to compute (whenever the method is possible to be applied)
explicitly all LRS spacetimes which admit proper RCs and MCs as well as the
collineation vector itself. The class (\ref{sx1.1}) is the richer in
admitting these higher collineations and for this reason we have considered
various examples which show on the one hand the usefulness and the
generality of the results and on the other the way one should follow in the
exploitation of Tables 1,2. Perhaps we should remark that we have obtained
without any effort all homogeneous LRS spacetimes and also all static
spherically symmetric spacetimes admitting RCs and MCs a subject which has
been considered many times in the literature the first part of it answered
completely only very recently \cite{Ziad}.

The physical applications we have considered are the most immediate and the simplest
ones and they do not really show the importance of the results or their
power, which could be used in many ways and directions. This will be the subject of a future work. 
\\\\\\
{\Large \textbf{Appendix 1}}\\\\
The KVs which span the $G_4$ are \cite{Apostol-Tsamp5}:

\[
{\bf K}_1=\partial _x,\qquad {\bf K}_2=\partial _z 
\]
\begin{equation}
{\bf K}_3=f(y,k)\cos z\partial _x+\sin z\partial _y+\left[ \ln \Sigma
(y,k)\right] _{,y}\cos z\partial _z  \label{A1}
\end{equation}
\[
{\bf K}_4=f(y,k)\sin z\partial _x-\cos z\partial _y+\left[ \ln \Sigma
(y,k)_{,y}\right] \sin z\partial _z 
\]
where: 
\begin{equation}
f(y,k)=\Lambda (y,k)\left[ \ln \frac{\Lambda (y,k)}{\Sigma (y,k)}\right]
_{,y}  \label{A2}
\end{equation}
They have the Lie brackets:

\begin{eqnarray}
\lbrack {\bf K}_1,{\bf K}_2] &=&0,\qquad [{\bf K}_1,{\bf K}_3]=0,\qquad [%
{\bf K}_1,{\bf K}_4]=0,\qquad [{\bf K}_2,{\bf K}_3]=-{\bf K}_4  \nonumber \\
&&  \label{A3} \\
\lbrack {\bf K}_2,{\bf K}_4] &=&{\bf K}_3,\qquad [{\bf K}_3,{\bf K}_4]=-k%
{\bf K}_2+2(1-k^2){\bf K}_1.  \nonumber
\end{eqnarray}
Let us assume that the ''metrics'' (\ref{sx3.1}) admit exactly one more KV,
the ${\bf X}_I$ say, which together with the $G_4$, generates an isometry
group $G_5$. Considering the commutator of ${\bf X}_I$ with the KVs ${\bf K}%
_1,{\bf K}_2,{\bf K}_3,{\bf K}_4$ and using Jacobi identities and Killing
equations we compute easily the form of ${\bf X}_I$ and the ''metric'' (\ref
{sx3.1}). The cases with 6 or 7 KVs are excluded because they lead to
conformally flat ''metrics''.
\\\\\\
{\Large\textbf{Appendix 2}}\\\\
We compute the Weyl tensor for the ''generic'' metric (\ref{sx3.1}) and find
that conformal flatness implies the condition: 
\begin{equation}
K_1=cK_2  \label{B1}
\end{equation}
\begin{equation}
c^2\varepsilon _2\Lambda ^{\prime 2}+\varepsilon _3\Sigma \Sigma ^{\prime
\prime }=0  \label{B2}
\end{equation}
where $c$ is a constant of integration. Hence $c^2=1$ and $\varepsilon
_2=k\varepsilon _3$ from which it turns out that $c=k$ i.e. the essential
constant is the curvature of the 2-dimensional space.

In terms of the ''generic'' metric the condition (A2) means that the
3-dimensional ''metric'' $K_{\alpha \beta }$ has either Euclidean or
Lorentzian signature and the overall signature of the ''generic'' metric is
Euclidean, Lorentzian and $(++--)$.

\end{document}